# COMPARISON OF PROCESS OF DIFFUSION OF INTERSTITIAL OXYGEN ATOMS AND INTERSTITIAL HYDROGEN MOLECULES IN SILICON AND GERMANIUM CRYSTALS: QUANTUMCHEMICAL SIMULATION

Vasilii Gusakov

Joint Institute of Solid State and Semiconductor Physics, P. Brovka str. 17, 2200 72 Minsk, Belarus

The theoretical analysis of the process of diffusion of interstitial oxygen atoms and hydrogen molecules in silicon and germanium crystals has been performed. The calculated values of the activation energy and pre-exponential factor for an interstitial oxygen atom $E(Si) = 2.59$ eV, $E(Ge) = 2.05$ eV, $D(Si) = 0.28$ cm$^2$ s$^{-1}$, $D(Ge) = 0.39$ cm$^2$ s$^{-1}$ and interstitial hydrogen molecule $E(Si) = 0.79 - 0.83$ eV, $E(Ge) = 0.58 - 0.62$ eV $D(Si) = 7.4 \cdot 10^{-4}$ cm$^2$ s$^{-1}$, $D(Ge) = 6.5 \cdot 10^{-4}$ cm$^2$ s$^{-1}$ are in an excellent agreement with experimental ones and for the first time describe perfectly an experimental temperature dependence of an interstitial oxygen atom and hydrogen molecules diffusion constant in Si and Ge crystals. It is shown, that for a case of impurity atom with a strong interaction with a lattice (interstitial oxygen atom) process of diffusion has a cooperative nature - the activation energy and pre-exponential are controlled by the optimum position of three nearest lattice atoms. For a case of extended defect with a weak interaction (an interstitial hydrogen molecule) process of diffusion is determined by the activation barrier subjected to fluctuations connected with rotation of a hydrogen molecule. The effect of hydrostatic pressure on the process of diffusion is discussed also.

## INTRODUCTION

Development of theoretical methods of determining the diffusivity of atoms in crystals is of great interest not only from a fundamental – understanding microscopic process of diffusion, but also from a practical point of view. The reasoning is that practically all technological processes of microelectronics in some way or another are connected to diffusion of atoms

(defects) in crystals. Moreover, the diffusion in crystals (nanostructures) occurs very often under extreme conditions (very high temperatures, fields of stress, interfacial regions, nanomaterials etc) and that essentially impedes, makes expensive or even impossible an experimental research. However until now there have been many obscure questions related to the microscopic mechanism of diffusion in crystals, whenever migration of an impurity atom involves the breaking and forming of covalent bonds. Recently I advanced the general method of calculation of coefficient of diffusion of interstitial atom of oxygen in silicon and germanium crystals [1, 2]. Theoretical analysis performed in [1, 2] enables to calculate the diffusion coefficient-the pre-exponential factor and activation energy both at normal pressure and at a hydrostatic compression of a crystal lattice. The calculated values of coefficient of diffusion are in very good agreement with experimental results [1, 2]. An interstitial oxygen atom is strongly interacting atom with a crystal lattice. In this connection it is obviously important to consider other limiting case - diffusion of a particle weakly interacting with a crystal lattice and comparison of features of diffusion of strongly and weakly interacting particles with a crystal lattice. As a characteristic example in this article the comparative analysis of process of diffusion of interstitial oxygen atom (particle strongly interacting with a crystal lattice) and the interstitial hydrogen molecule(particle weakly interacting with a crystal lattice) in silicon (germanium) crystals under normal and hydrostatic pressure (HP) is carried out. To the best of my knowledge, no effects of HP on the $H_2$ diffusivity have been considered yet.

**THEORETICAL APPROACH**

Let us consider the physical parameters determining the diffusion process of an atom in a crystal. Modeling by the method of random walk results in the following general expression for the diffusion constant:

$$D = \frac{d^2 N_{et}}{2d_s} \Gamma \qquad (1)$$

where $d$ is diffusion jump distance, $N_{et}$ is the number of the equivalent trajectories leaving the starting point, $d_s$ is the dimension of space (in our case $d_s=3$), $\Gamma$ is the average frequency of jumps on the distance $d$. In the case of a system consisting of $N$ atoms, using the reaction-rate theory [3], the value of $\Gamma$ may be written in the following form

$$\Gamma = \frac{1}{2\pi} \frac{\prod_{i=0}^{N} \lambda_i^{(o)}}{\prod_{i=1}^{N} \lambda_i^{(b)}} \exp\left(-\frac{\Delta E_a}{k_B T}\right), \qquad (2)$$

where $\Delta E_a$ is the adiabatic potential energy difference between the saddle point and the stable one, $\lambda_i^{2(o,b)}$ are the eigenvalues of the matrix (with respect to mass-weighted internal coordinates) $K_{ij} = \partial^2 U_{eff} / \partial f_i \partial f_j$, $U_{eff}(f_1,...,f_m)$ denotes the potential as a function of the internal degrees of freedom. The indices (b) and (o) indicate that the corresponding quantities are evaluated at the saddle point and local minimum, respectively. Thus, the diffusion constant $D$ is determined by the following diffusion parameters: the length of diffusion jumps ($d$), the diffusion barrier ($\Delta E_a$), the number of equivalent ways leaving the starting point of diffusion jumps ($N_{et}$) and the eigenvalues matrix ($\lambda_i^{2(o,b)}$). The calculation of diffusion parameters was performed in a cluster approximation. For comparison with the previous calculations different methods such as empirical potential (MM2), semiempirical (AM1, PM3, PM5) and ab-initio (RHF, LDA) have been used for the calculation of the cluster total energy. Depending on the method of total energy calculation the cluster size was varied from 17 Si atoms (ab-initio methods) up to $10^3$ Si atoms (semiempirical and empirical potential methods).

**RESULTS AND DISCUSSION**

The starting and the final points of the diffusion jump correspond to the equilibrium configuration of an interstitial oxygen and interstitial hydrogen molecule (H$_2$) in a crystal. Individual oxygen atom occupies interstitial bond-center (BC) position in silicon and to diffuse by jumping between the neighboring BC sites. Hence the starting and the final points of the diffusion jump correspond to the equilibrium BC configuration of an interstitial oxygen atom (O$_i$) in silicon. The calculated equilibrium configuration of O$_i$ and the local vibration frequency of the asymmetric stretching mode (B$_1$) takes the following values: $d_{Si-O}$ =1.63 Å (RHF, 6-31G$^{**}$), 1.582 Å (MM2), 1.61 Å (PM5), $\angle_{Si-O-Si}$ =161.6º (RHF, 6-31G$^{**}$), 167.3º (MM2), 171º (PM5), $\nu_O$ = 1214 cm$^{-1}$ (RHF, 6-31G$^{**}$), 1091 cm$^{-1}$ (AM1), 1078 cm$^{-1}$ (LDA) and are in a good agreement with experimental [4, 5] and recently calculated ones [6, 7]. The calculated value of the potential barrier for the rotation of O$_i$ around Si-Si axis equals $\Delta E_\varphi \leq 20$ meV (PM5). As $\Delta E_\varphi$ is much less than $k_B T$ (at diffusion temperatures) an interstitial oxygen atom jumps on any of six nearest Si-Si bonds (Fig. 1.) and, hence, in Formula (1) the parameters $N_{ef}$= 6 and $d$=0.19 nm. The calculated equilibrium configuration of H$_2$ corresponds to position of the centre of mass of a molecule in the tetrahedral interstitial (T) site of crystal. The calculated equilibrium configuration of H$_2$ and the stretch vibrational frequencies takes the following values: $d_{H-H}$ (Ge) =0.0753 nm, $d_{H-H}$ (Si) =0.074 nm, $\nu$(Si) = 3413 cm$^{-1}$, $\nu$(Ge) = 3586 cm$^{-1}$ (PM3, PM5) and are in a good agreement with recently calculated ones [8]. The Mulliken population analysis provides a qualitative value for the effective charge associated with each atom. Both H atoms in H$_2$ carry a small positive charge (Q(H$_2$) ~+0.06e). Some electron density is transferred from the H$_2$ molecule to its Si neighbors. The calculated value of the potential barrier for the rotation of H$_2$ equals $\Delta E_\phi \leq 6$ meV (PM5). The barrier $\Delta E_\phi$ for rotation of the H–H dumbbell around its CM is much less than $k_B T$, implying a nearly-free rotator at the T site. Owing to thermal fluctuations the CM moves randomly in the volume limited by isoenergetic surface with the maximal displacement $\Delta r_{max} \approx 0.025$ nm. The value $\Delta r_{max}$ is in a good agreement with the results of modeling by a molecular dynamic method [8]. Since the isoenergetic surface is anisotropic (calculated normalized diagonal matrix of the second derivatives in T configurations has the following value $\partial^2_{xx} E = \|1, 0.79, 0.7\|$), the diffusion transition occurs in a direction of maximal displacement $H_2$ between the nearest T configurations along the trigonal axis from T to hexagonal to T sites (T-H-T path). In this case the number of equivalent trajectories $N_{et} = 1$ and the calculated distance of diffusion transition has the following value $d = 0.115 - 0.12$ nm.

To determine the activation barrier of diffusion the simulation of probable diffusion trajectories was performed. The oxygen atom (hydrogen molecule) was displaced from the equilibrium configuration along a trajectory in the direction to a new equilibrium position (nearest Si-Si bonds for the case of $O_i$ or T site for the case of $H_2$ molecule). Owing to thermal fluctuations crystal lattice atoms may occupy any permissible positions at the given position of $O_i$ or $H_2$ but all of these configurations differ in the total energy of a crystal. Since the diffusion constant exponentially depends on the diffusion barrier (2) we should pick the minimal value of $\Delta E_a$

$$\Delta E_a = \min[E_{cl}(S_G) - E_{cl}(O)], \tag{3}$$

where $E_{cl}(S_G)$ and $\Delta E_{cl}(O)$ are the total cluster energy on the surface $S_G$ (on this surface energy has a local maximum) and in local minimum O (equilibrium configuration), respectively. It means that for determining of the minimal value of $\Delta E_a$ we should minimize total energy of cluster as a function of atoms coordinates of crystal lattice. Along the given trajectory the minimal total cluster energy has been calculated and among the set of calculated trajectories the extreme trajectory satisfying condition (3) was selected. The simulation has revealed an important fact for understanding of the diffusion process. $E_{cl}(S_G)$ and hence $\Delta E_a$ depends on the number of the crystal lattice atoms ($n$) nearest to diffused particle ($O_i$ or $H_2$) involved in the minimization of the cluster total energy. The activation barrier $\Delta E_a(n)$ decreases with increase of the number of crystal lattice atoms participating in minimization (Fig 1.). Therefore, first of all, we should determine how many of the nearest crystal lattice atoms are involved in the diffusion process. A diffused particle can overcome a barrier at any optimum configuration of crystal lattice atoms. However the relative number of particles diffused at the given optimum configuration of crystal lattice atoms is proportional to the product of probabilities of occurrence of the optimum configuration $P_{occ}$ and probability of diffusion jump $P_{dj}$ ($P_{dj} \propto \exp(-\Delta E_a(n)/k_B T)$). The probability of occurrence of an optimum configuration out of $n$ atoms have been calculated on the basis of geometrical definition of probability (the problem of a random collisions) and in this case the product $P_{occ}P_{dj}$ may be written as

$$P_{occ}P_{dj} \propto n \left( \frac{\Delta \tau(n)}{\tau(n)} \right)^{n-1} \exp\left( -\frac{\Delta E_a(n)}{k_B T} \right), \qquad (4)$$

where $n$ is the number of atoms in the optimum configuration, $\tau(n)$ and $\Delta\tau(n)$ are the period of formation and lifetime of the given optimum configuration, respectively, $\Delta E_a(n)$ is the diffusion barrier. Usually $\Delta\tau(n)/\tau(n)$ is much less than one [3]. The dependence $P_{occ}P_{dj}$ as function of $n$ Si atoms involved in the minimization is depicted in the insets of Fig. 1. Calculations have shown, that $P_{occ}P_{dj}$ for interstitial oxygen atom has a sharp maximum at $n=3$ which more than an order of magnitude exceeds $P(n)$ for $n=2, 4, 5…$ Hence, essentially all $O_i$ atoms overcome the diffusion barrier when three nearest Si atoms (Si atoms connected to $O_i$ atom before and after diffusion jump Si(1) - $O_{initial}$ – Si(2) – $O_{final}$ – Si(3)) are in the optimum configuration. The result obtained suggests that for a case of impurity atom with a strong interaction with a lattice (interstitial oxygen atom) process of diffusion has a cooperative nature - the activation energy and pre-exponential are controlled by the optimum position of three nearest lattice atoms and the diffusion parameters should be calculated for the given configuration. For interstitial hydrogen molecules $P_{occ}P_{dj}$ (n) is a rapidly decreasing function. It means that practically all $H_2$ molecules diffuse without correlation with the position of the nearest atoms of crystal lattice and in this case the diffusion transition is not a cooperative transition.

Let's consider variation of total cluster energy lengthways of diffusion trajectories. A number of the dependences calculated near to a maximum of a potential energy are presented in Fig. 2. For interstitial hydrogen molecule the maximum of a potential barrier is located symmetrically relative initial and final points of diffusion transition of $H_2$ and potential function of energy is a saddle surfaces. As orientation of $H_2$ molecule in a saddle point is arbitrary, the activation barrier is a stochastic function of orientation of $H_2$ molecule and the process of diffusion of a rotating molecule should be considered as process of diffusion in a fluctuating potential [9]. The activation barrier was evaluated for three orientations of $H_2$ molecule in a saddle point (the axis of a $H_2$ molecule is directed lengthways x. y, z coordinate axes where the axis x was perpendicular planes H). For a molecule of hydrogen the diffusion transition is not cooperative, and as consequence, the maximum of a potential barrier is located symmetrically concerning initial and final points of direct and reverse diffusion transition. In the case of interstitial oxygen atom the variation of total cluster energy lengthways of diffusion trajectories has essentially other nature (Fig. 2.) For the extreme trajectory (3) the maximum of the potential energy (a saddle point of Oi migration) is displaced from the midpoint of the path in a direction to a final point of diffusion transition and a direct trajectory do not coincide with a return trajectory. The displacement of a maximum of potential energy is caused by cooperative nature of diffusion transition - correlated displacement of three atoms - Si(1) - $O_{initial}$ - Si(2) - $O_{final}$ - Si(3) occurs, and the displacement is far more for Ge crystals. In a vicinity of a maximum the abrupt variation of potential energy is observed. Similar dependence of potential energy lengthways of diffusion trajectories is characteristic for atoms strongly interacting with a crystal lattice. Really, abrupt variation of potential energy in vicinity of a maximum is caused by reconfiguration of an electronic subsystem of a crystal. In the course of transition of $O_i$ atom from one equilibrium configuration in another the breaking of old and formation of new covalent Si-O bonds takes place. The process of reconfiguration of an electronic subsystem will occur in the case when oxygen and neighboring silicon atoms owing to thermal fluctuations get in the region of configuration space G (bounded by the critical surface $S_G$) where the electronic reconfiguration leads to lowering of the crystal total energy. Matrix $\lambda_i^{2(o,b)}$ necessary for the calculation of the pre-exponential factor $D_0$ was evaluated as follows. At the equilibrium configuration of interstitial oxygen $O_i$ or interstitial hydrogen molecule $H_2$ and at the intersection point of the extreme trajectory of $O_i$ or $H_2$ with surface $S_G$ the square-law interpolation of the potential energy $U_{eff}(f_1,...,f_m)$ ($f_1$... $f_m$ are coordinates of $O_i$ or $H_2$ and nearest Si atoms) has been constructed and $\lambda_i^{2(o,b)}$ has been obtained at once by diagonalization of $K_{ij}$. Calculated values of pre-exponential factor and activation barrier for diffusion of interstitial oxygen atom and interstitial hydrogen molecules are presented in Table I and Fig. 3. On Figure 3 one can see the excellent agreement between the calculated and experimental temperature dependences of the diffusion coefficient in all temperature range $T$=350 – 1200 °C for $O_i$ and T= 40 – 200 °C for $H_2$.

For better understanding of $O_i$ and $H_2$ diffusion the influence of compressive hydrostatic pressure (HP) on the diffusion coefficient has been evaluated. This is particularly interesting as high HP has been found to enhance strongly the oxygen agglomeration at elevated temperatures [10 – 13]. Recently [1, 2, 14] I have shown that under HP the activation barrier of diffusion of interstitial oxygen in Si crystals is decreased and that is the reason to enhance strongly the oxygen agglomeration at elevated temperatures. Modelling of influence of HP upon the diffusion of interstitial hydrogen molecule was carried out as in [14]. The cluster has been conventionally divided into internal ($R < R_0$) and external parts ($R>R_0$). The internal part includes diffused particle and is selected in such a manner that the increase in $R_0$ does not result in essential change of the equilibrium structure of $O_i$ or $H_2$ defect at pressure $P$=0 (usually $R_o$ equals 0.5 – 0.7 nm). The pressure has been modeled by replacement of the equilibrium length of bonds in the external part of the cluster with the length of Si-Si bonds that are characteristic (calculated from experimental value of compressibility modulus) for the given pressure. Upon minimization of the cluster total energy, the lengths of bonds at $R>R_0$ did not vary, and the minimization was carried out on the coordinates of $O_i$ or $H_2$ and crystal lattice atoms being in the internal part of the cluster. The further evaluation of $\Delta E_a(P)$ was carried out similarly to the case $P=0$. Calculations have revealed that HP leads to variation of the activation barrier. Linear interpolation of $\Delta E_a(P)$ dependence

$$\Delta E_a(P)/\Delta E_a(0) = 1 + \gamma \cdot P \qquad (5)$$

result in the following value of $\gamma(O_i) = -(1.69 \pm 0.1)\ 10^{-11}\ Pa^{-1}$ and $\gamma(H_2) = (0.92 \pm 0.35)\ 10^{-11}\ Pa^{-1}$, $P$ is the HP. The calculated pressure dependence of the $O_i$ diffusivity (without any adjustable parameters) corresponds well to an enhanced growth of the oxygen-related thermal donors (TDs) observed experimentally [1, 2, 14]. It is interesting to note that HP decreases the activation barrier for diffusion of interstitial oxygen ($\gamma(O_i) < 0$) also increases activation barrier for diffusion of an interstitial hydrogen molecule ($\gamma(H_2) > 0$). Qualitatively it is possible to explain the result obtained as follows. In a case of interstitial oxygen atom the value of activation barrier is determined by deformation of covalent Si-O bond in a saddle point. HP results in compression of a crystal and the distance between initial and final points of diffusion transition is decreased. Hence, in a saddle point the value of deformation of covalent Si-O bonds will be less and the activation barrier will decrease with increase of pressure. In the case of interstitial hydrogen molecule the activation barrier is determined by interaction of $H_2$ molecule and the nearest Si atoms in the saddle point. Under HP the distance between Si atoms in a saddle point is decreased, overlapping of molecular orbitals of hydrogen molecules and Si atoms are increased and, hence, the growth of activation barrier is observed. The calculated value $\gamma(H_2)$ has a sufficiently big average mistake that points to a possible nonlinear dependence of $\Delta E_a(P)$. Moreover, as preliminary calculations have shown in case of diffusion of interstitial $H_2$ in Ge and Si crystals the pre-exponential factor is a function of HP and that demands more detailed calculations of coefficient of diffusion of $H_2$ under hydrostatic pressure. These calculations are in progress.

**SUMMARY**

In summary, a theoretical modeling and comparative analysis of the interstitial oxygen and interstitial hydrogen molecule diffusivity in silicon and germanium crystals at normal and hydrostatic pressure has been presented. On the basis of the results obtained it is possible to draw the following conclusions. The process of diffusion of strongly interacting particle (interstitial oxygen atom) with crystal lattice is a cooperative process. Three nearest crystal lattice Si (Ge) atoms are involved in an elementary oxygen jump from a bond-center site to another bond-center site along a path in the (110) plane. It is precisely their optimum position (corresponding to a local minimum of the crystal total energy) which determines the value of the diffusion parameters of an interstitial oxygen atom in silicon and germanium. For the case of weakly interacting particle with crystal lattice (interstitial hydrogen molecules) the process of diffusion is not correlated with the position of crystal lattice atoms. But in the case of $H_2$ the diffusion parameters is subjected to fluctuations connected with rotation of a hydrogen molecule. The theoretically determined values of the diffusion potential barrier and pre-exponential factor are in excellent agreement with experimental ones and describe very well the experimental temperature dependence of diffusion constant in Si crystals. Hydrostatic pressure gives rise to the decrease of the diffusion potential barrier for interstitial oxygen and increases the diffusion potential barrier for interstitial hydrogen molecule in Si crystals  Such a pressure dependence of $O_i$ diffusivity appears most likely to be responsible for the HP enhancement in generation of the oxygen-related thermal donors.

**ACKNOWLEDGMENT**

The author is grateful to the CADRES, INTAS and the State Committee for Science and Technology and Fund for Fundamental Research of the Republic of Belarus (grants 01-0468, 03-50-4529 ) for financial supports.

Table I. Calculated values of diffusion parameter for interstitial oxygen and interstitial hydrogen molecules in silicon and germanium crystals

| | Si | | | | Ge | | | |
|---|---|---|---|---|---|---|---|---|
| | Theory | | Experiment [15, 16, 17] | | Theory | | Experiment [18] | |
| | $O_i$ | $H_2$ | $O_i$ | $H_2$ | $O_i$ | $H_2$ | $O_i$ | $H_2$ |
| $\Delta E_a$, eV | 2.59 | 0.79 – 0.83 | 2.53 | 0.78±0.05 | 2.05 | 0.58 – 0.63 | 2.076 | - |
| $D_0$, cm$^2$ s$^{-1}$ | 0.28 | 7.5·10$^{-4}$ | 0.13 | (2.6±1.5)·10$^{-4}$ | 0.39 | 6.5·10$^{-4}$ | 0.4 | - |

**CAPTURES ON FIGURES**

Figure 1. Diffusion barrier $\Delta E_a(n)$ as a function of the number of Si atoms involved in minimization of the total cluster energy. In the inset the probability $P_{occ}P_{dj}$ of occurrence of an optimum configuration out of *n* atoms is presented *($\Delta\tau(n)/\tau(n)$=0.01)*.

Figure 2. Variation of total cluster energy lengthways of diffusion trajectories.

Figure 3. Temperature dependence of diffusion constant of interstitial oxygen atom and interstitial hydrogen molecules in silicon. Points – experiment [15, 17], line – theory.

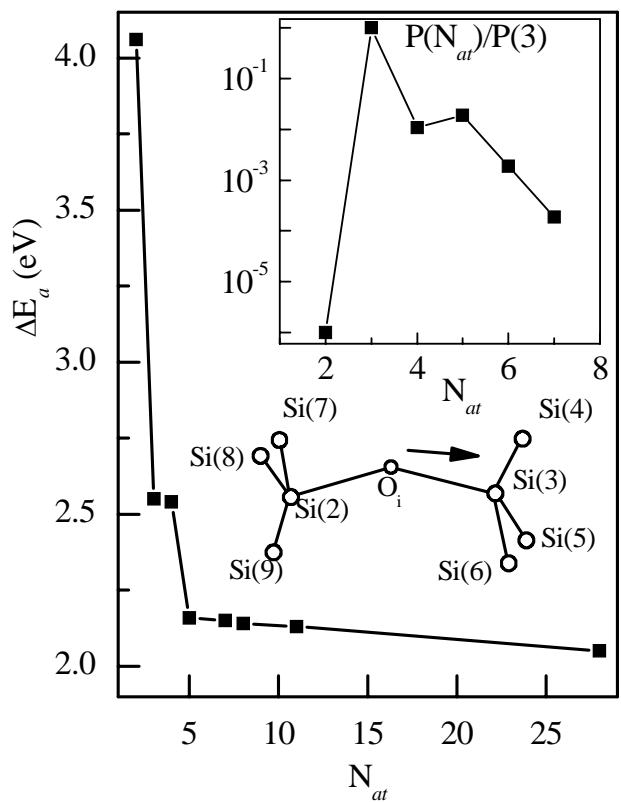
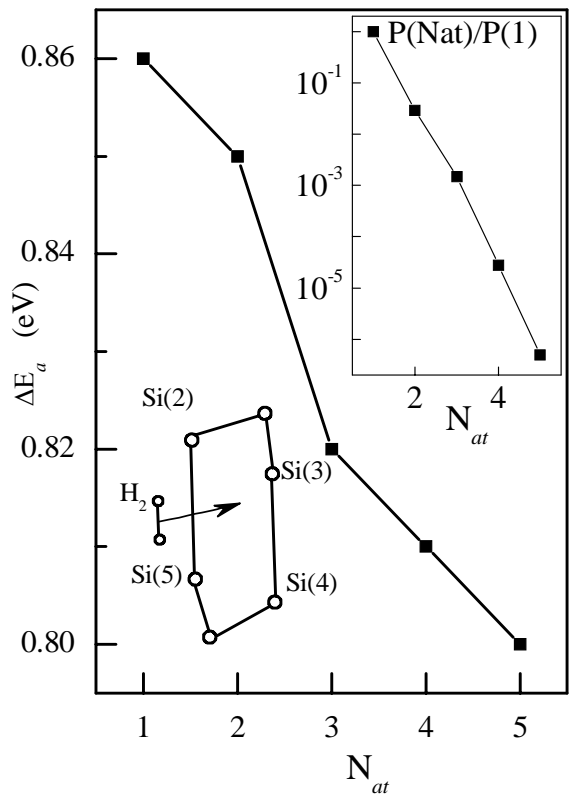

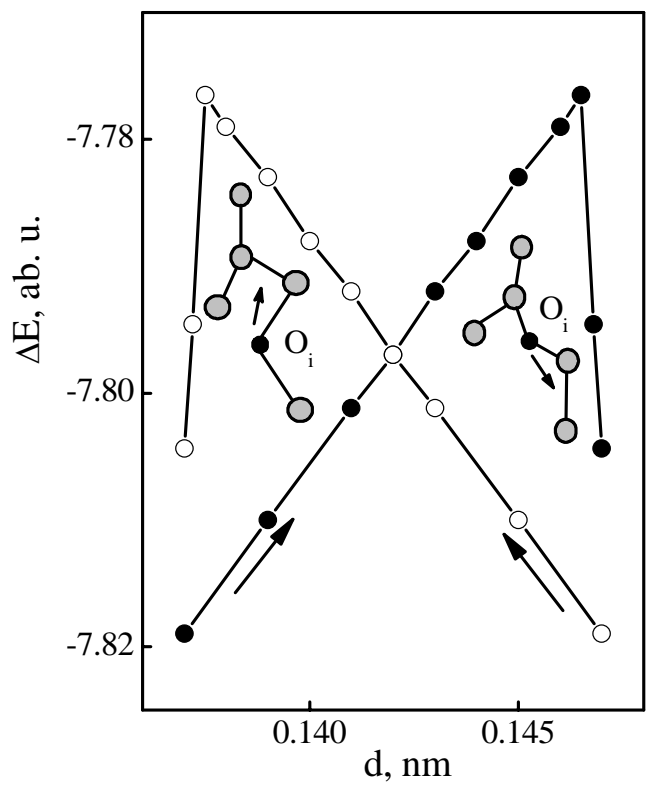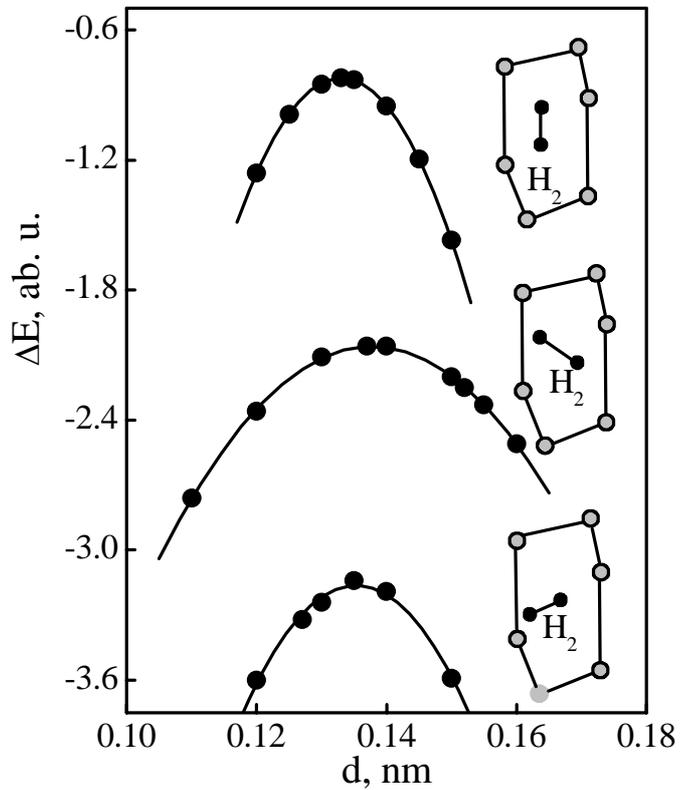

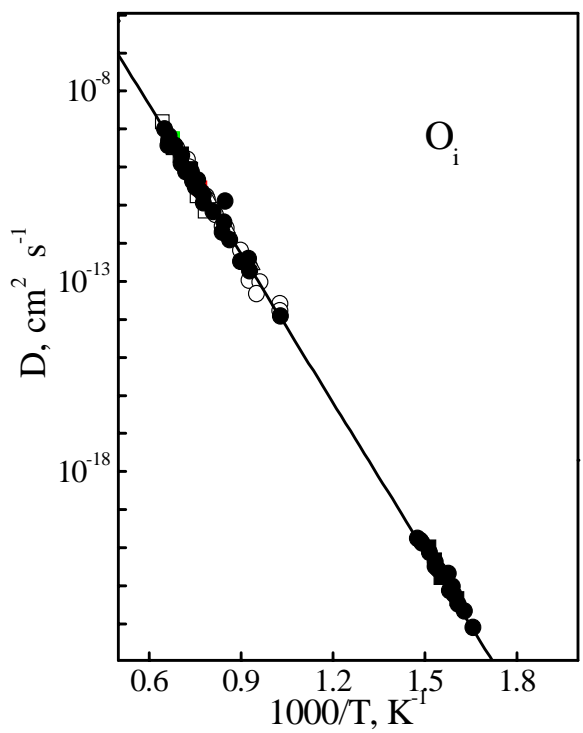 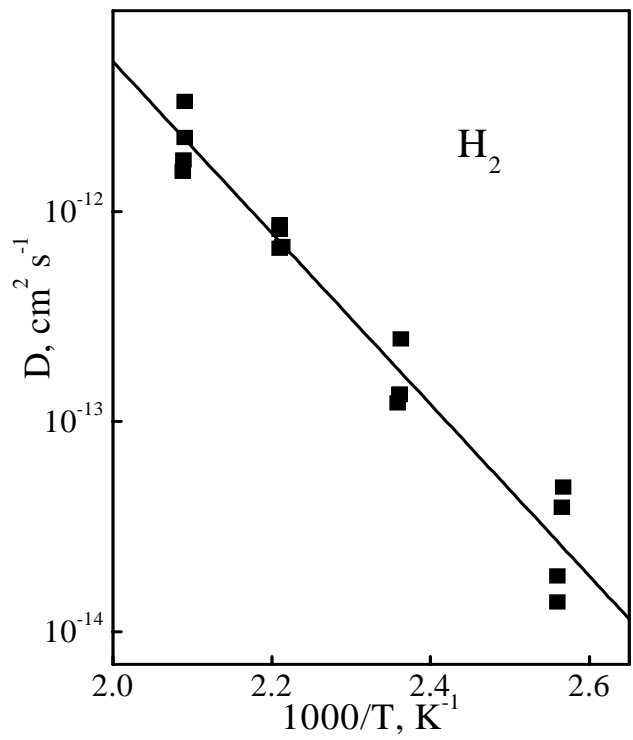

3